\documentclass[aip,apl,showpacs,reprint]{revtex4-1}

\usepackage{times}
\usepackage{amsmath,amssymb}           
\usepackage{graphicx}                   
\usepackage{latexsym}
\usepackage{sidecap}

\begin{document}

\title{Ultra small mode volume defect cavities in spatially ordered and disordered metamaterials}
\author{Nad\`{e}ge Kaina}
\author{Fabrice Lemoult}
\author{Mathias Fink}
\author{Geoffroy Lerosey}
\email[]{geoffroy.lerosey@espci.fr}
\affiliation{Institut Langevin, ESPCI ParisTech \& CNRS UMR 7587, 1 rue Jussieu, 75005 Paris, France}
\date{\today}

\begin{abstract}
In this letter we study metamaterials made out of resonant electric wires arranged on a spatial scale much smaller than the free space wavelength and we show that they present a hybridization band that is insensible to positional disorder. We experimentally demonstrate defect cavities in disordered and ordered samples and prove that, analogous to those designed in photonic crystals, those cavities can present very high quality factors. In addition we show that they display mode volumes much smaller than a wavelength cube, owing to the deep subwavelength nature of the unit cell. We underline that this type of structure can be shrunk down to a period close of a few skin depth. Our approach paves the way towards the confinement and manipulation of waves at deep subwavelength scales in both ordered and disordered metamaterials.

\end{abstract}

\pacs{}

\maketitle

Photonic crystals (PC) are periodic composite materials typically scaled at the wavelength which
exhibit band gaps due to their translational symmetries\cite{PhysRevB.19.5057,Yablonovitch,
PhysRevLett.58.2486,PhysRevLett.65.3152,Joannopoulos,meade:495,Yablonovitch:93}. As these 
band gaps result from interferences between distinct paths followed by a propagating wave, 
local modifications of the medium does not affect its global properties and can create defect cavities 
in which the waves are trapped\cite{PhysRevLett.67.2295}. Conversely metamaterials, 
which are organized at a scale much smaller than the wavelength, are usually studied under an effective medium approach and designed to obtain macroscopic effective properties such as 
negative indices\cite{pendrySuperlens,Veselago,PRLSmith2000,valentine2008,FishnetLalanne}.
In a recent paper however\cite{NatPhys}, we have shown that it is possible to merge the advantages of 
metamaterials and PC, hence allowing the control of waves on a deep subwavelength scale, using locally
resonant media. We have proved that they present a so called hybridization band gap\cite{PSheng1991,PhysRevB.65.064307,LeroyEPJE2009} which can be explained by
the far field coupling between the resonance of the unit cells and the incoming plane waves. This has allowed us to  demonstrate waveguides, cavities, benders or splitters, of dimensions very small compared to the wavelength.  

In this letter we go beyond this proof of principle work. We experimentally study locally resonant metamaterials in the microwave domain and we exploit the hybridization bang gap that they present. We prove that the latter is due to the resonant nature of their constitutive unit cell rather than to symmetries by characterizing both spatially ordered and disordered metamaterials.
Then we demonstrate that a local modification of the medium permits to create ultra small mode volume defect cavities in both ordered and disordered samples. We show that those cavities exhibit at the same time high quality factors, around half a thousand, and ultra low volume modes, less than $\lambda^{\mathrm{3}}$/10000. This results in extremely high Purcell factors.  We underline that interestingly, shrinking the medium in its transverse dimension decreases at the same time the losses and the mode volume of the cavity, which results in never achieved Purcell factors \cite{purcell}. We finally discuss the limits of the idea and notably its transposition to higher frequency parts of the spectrum.

\begin{figure} [!h]
\begin{center}
\includegraphics[width=0.9\columnwidth]{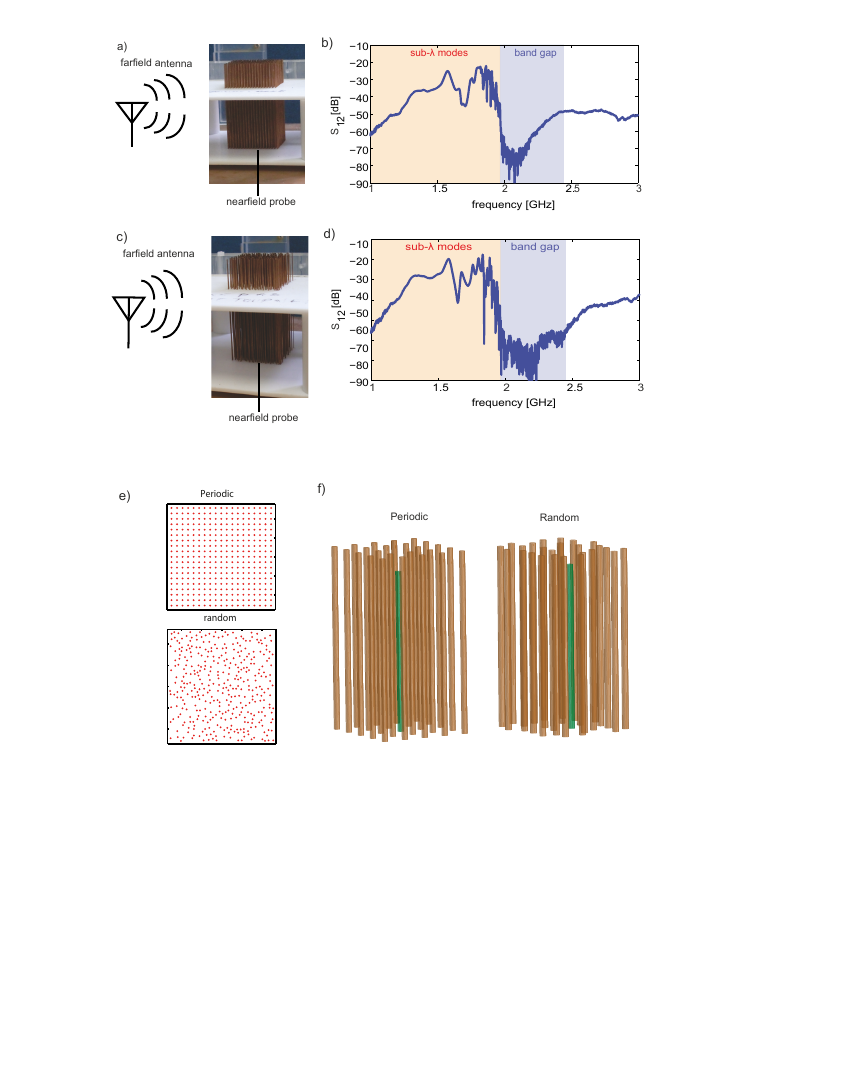}
\caption{Experimental samples: a periodic (a) and a spatially random (c) wire medium. (b-d) Measure of the hybridization band gap of both samples. (e) schemed top view of the samples under study, showing the spatial distribution of wires. (f) Point defect in the hybridization band gap for the periodic (left) and spatially random (right) sample. This point defect consists in introducing a shorter wire with a resonant frequency $f_1$ within the band gap of the metamaterial.} 
\label{Fig.1}
\end{center}
\end{figure}


In order to study experimentally the proposed concepts, we use a very 
academic subwavelength resonator, that is, a resonant metallic wire. Since 
it resonates along its long dimension, which we arbitrarly align with the
\textit{Oz} direction, the latter can 
be of deep subwavelength size in the \textit{xOy} transverse plane. In fact, the 
diameter of such a wire can even be decreased down to a few skin depths. Because 
we realize our samples manually, we opt for wires of length $L=7$ cm, which 
corresponds to a resonance around $f_{0}=2.15$ GHz, and of diameter $d=0.5$ 
mm. We start by measuring the properties of a periodic medium made out of 
$19\times19$ wires separated by a period $a=2$ mm (corresponding to $\lambda 
_{\mathrm{0}}$/70) maintained by a Teflon structure, and those of a 
randomly positioned one which contains the same average density of 
resonators (Fig 1.a and 1.c). In order to study the hybridization band gap 
that displays those media, we place azimuthally in the far field a vertically 
polarized antenna that is connected to one port of a network analyzer 
(Agilent N5230C PNA). We connect the second port to a handmade probe, a very 
electrically short monopole. This probe is placed in the near field of the 
arrays of wires, and is mainly sensitive to the vertical polarization of the 
electric field.

We measure the transmission coefficient between the two ports $S_{12}$. 
Because the monopole is very short, it is inefficient and reactive. This has 
two consequences: first, it is mainly sensitive to the evanescent field in 
the samples, and second, the transmission coefficient has to be understood 
as a relative measurement of the transmission. The 
corresponding $S_{12}$ are plotted as a function of the frequency in Figures 
1.b and 1.d. We know from our experiments of subwavelength focusing from the 
far field that such media support deep subwavelength modes that convert 
efficiently to the far field\cite{ScienceGeo,metalens,WRMLemoult1,WRMLemoult2}. 
Those modes are observable in the obtained curves as resonant peaks. We stress
here that the random medium of Figure 1.c also supports those collective modes,
although these contain a distribution of wavevectors as opposed to a single one
for the periodic medium. We have explained those modes using a Bloch formalism 
through rigorous calculations of the Maxwell's equations\cite{WRMLemoult1}. We have also adopted 
a more phenomenological formalism based on the idea of a hybridization between the continuum of
plane waves propagating in the medium and local oscillators, which are interfering either in phase
($f<f_0$, modes are allowed) or in anti-phase ($f>f_0$, the hybridization band gap is created)\cite{NatPhys}.
Indeed, the curves of the transmission coefficient between the two ports of 
the network analyzer show a wide dip that starts around the frequency of 2 GHz 
(Figs. 1.b and 1.d). It is clear that this band gap exists for both the 
periodic medium and the random medium, meaning that it is not due to the 
periodicity of the structures, but rather to the resonant nature of their unit 
cells.


Following the proposals of our recent paper\cite{NatPhys}, our aim is
now to demonstrate cavities in the investigated media. Evidently, since the structure of most 
metamaterials and in particular those studied here is deeply subwavelength, 
removing one unit cell (as implemented in PC) does not have much consequences, 
since within one period of the medium, no state can exist. However, a defect 
is simply created by detuning the resonant freqency of a single unit cell so that
it falls in the band gap created by the other wires. To that aim, we just tune the length of a single wire to a
length $L'=\alpha L$, where $\alpha $\textless 1 is correctly chosen.
Thus, we shift its resonance frequency upward of a factor $1/\alpha $.
We stress here that the existence of the defect cavity can be understood, despite 
the deep subwavelength organization of the crystal, only because the hybridization
band gap results from interference effects only\cite{NatPhys}. Indeed, near field interactions between unit cells would result in spatially extended states rather than point defect states. Moreover, we point out that 
embracing an effective medium approach, adding a defect would solely result in a reduced response of the system, which would hide the interesting phenomena studied here.

\begin{figure}[!h]
\begin{center}
\includegraphics[width=1\columnwidth]{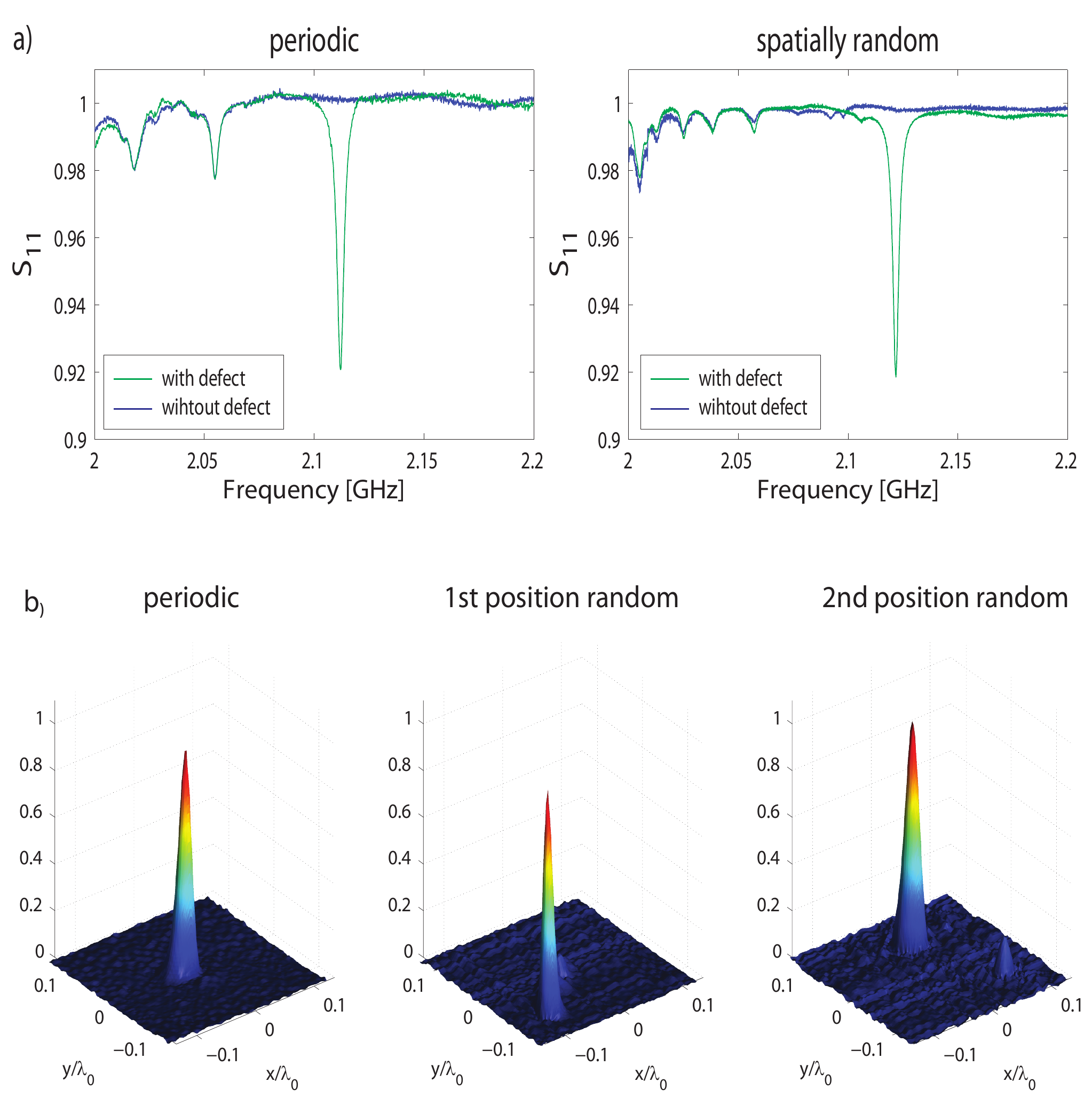}
\caption{(a) Measure of $S_{11}$ in the near field of the wire medium with or without the point defect in the periodic (left) and spatially random (right) sample. (b) Experimental measure of the confinement around the point defect in the periodic (left) and spatially random (center and right) sample.}
\label{Fig.2}
\end{center}
\end{figure}

We now test experimentally this concept using the fabricated samples. 
To do so, we first acquire with the network 
analyzer the reflection coefficient $S_{11}$ of the small monopole when it is 
placed in the near field of a chosen wire of length $L$.
This is realized for the two samples at frequencies ranging from 2 GHz to 2.2 GHz, 
and the result is plotted in blue in Figure 2.a for the periodic medium (right) and 
for the random medium (left). Below the frequency of 2.06 GHz, we can see dips 
in the reflection coefficient, meaning that some energy has been transmitted 
from the probe to the sample, or equivalently that the otherwise very 
inefficient monopole has been impedance matched by the medium due to the 
Purcell effect. These dips correspond to the higher frequency subwavelength 
modes supported by the samples.
Now we replace the chosen wire by a shorter one. The experiment is realized 
in both the periodic and the random media, for wire lengths 
that span the interval 65 mm to 70 mm, with a step of 0.5 mm. 
We obtained consistent results for all measurement and here we present those of the wire of length 
$L'=67$ mm, that is, one 3 mm shorter than those of the metamaterial. We plot in green curves the 
reflection coefficients obtained for the ordered and disordered samples in Figure 2.a. 
Clearly another dip has appeared for the two studied samples, and the latter 
corresponds to the defect created in the media, since it appears only after 
detuning the wire. Because the measured $|S_{11}|$ can be understood as an estimation of
the coupling between the probe and the medium, we conclude that at the resonance
frequency of the shortened wire, some energy is transmitted to the defect state.
We have consequently created a cavity for which we measure a quality factor around $Q=550$, 
which is more than two orders of magnitude higher than a single wire in vacuum. It is limited by 
the ohmic losses.

Of equal interest is the spatial extension of the cavity created by the 
defect. In order to obtain its estimation for a given modified wire of 
length $L'$, we use a 2D translation stage that permit us to scan the whole surface of 
the bottom of our samples. For each position, we measure the $S_{11}$ of the 
probe, meaning the coupling with the medium, at the resonance frequency of the defect. 
When the probe is located outside of the cavity, the reflection coefficient remains 
very close to one, while it decreases consequently above the defect cavity. The value 
of the $S_{11}$  at the resonance frequency of the defect state for each position in 
the near field of the sample gives hence an estimation of the spatial extension of 
the created cavity. In order to represent it graphically, and since the value of the 
reflection coefficient still has no exact meaning, we calculate the value \textbar 1-$S_{11}$\textbar  
as a function of the position in the \textit{xOy} plane. We map it for one position 
in the periodic medium and two positions in the random one (Fig.2b).

\begin{figure} [!h]
\begin{center}
\includegraphics[width=0.9\columnwidth]{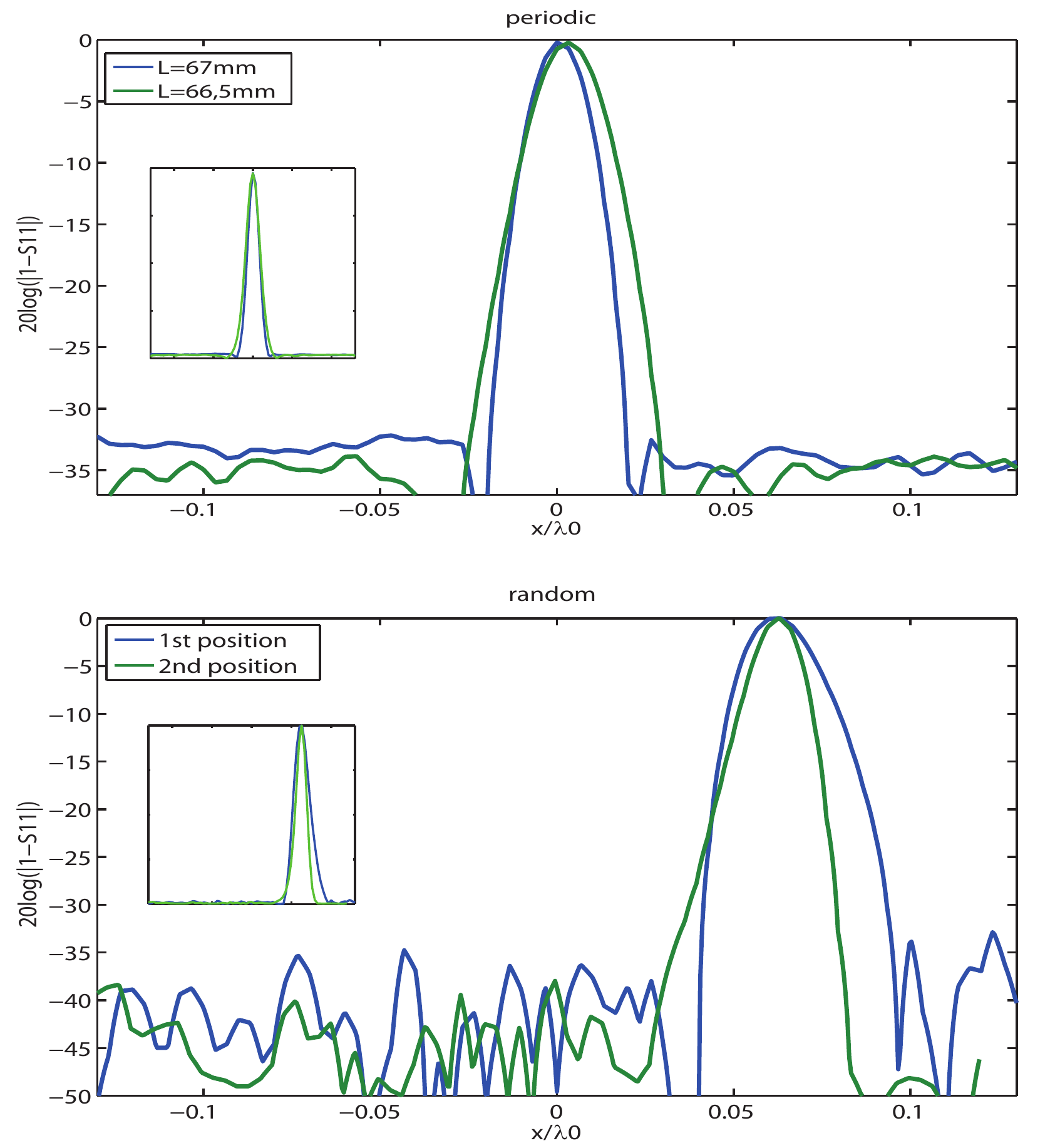}
\caption{Profile of the $|1-S_{11}|$ in logaritmic scale for two defect frequencies 
in the periodic array (up) and two positions in the random array (down). In inset are
 shown the corresponding linear profiles} 
\label{Fig.3}
\end{center}
\end{figure}

From Figure 2, it appears that the cavity width roughly measures one period of 
the medium in the transverse dimensions, that is, around $\lambda $/70 at 
its resonance frequency. This typical size can vary in the random medium, 
since the distance between neighbors changes with the position in the 
medium, but it remains of the same order. Given the fact that the mode in 
the cavity is transverse electromagnetic, the energy density is constant 
along the \textit{Oz} axis. This indicates that the total mode volume of the cavity is 
of the order of ($\lambda $/70)$^{\mathrm{2}}$.( $\lambda $/2), that is 
around $\lambda^{\mathrm{3}}$/10000. Indeed, not only is the period of 
such media much smaller than that of conventional photonic or phononic 
crystals, but their attenuation distance is also much smaller in the band 
gaps (Fig.1 b-d). To highlight this effect, we represent the profile of 
the $|1-S_{11}|$ along one dimension of the \textit{xOy} plane for both
defects in the periodic  and random media. From figure 3, we measure attenuation
coefficients of the order of 25dB per period in the periodic crystal as
well as in the random medium. In the latter, this coefficient slightly depends
on the position of the defect and on which transverse dimension we measure the profile,
since it changes the distance to the first neighboring wires.  

Naturally, the spatial extension of the field in the defect cavity has to be 
compared to that created around a resonant wire alone, since the latter already confines the 
electromagnetic field onto subwavelength dimensions. We have estimated that the mode
volume of such a resonator is 5 to 6 times larger than when it is placed in the
hybridization band gap material. This ratio is not huge, but we have already proven
that the quality factor of the cavity has been increased in the metamaterial by a factor
larger than a hundred. Overall, this results in a drastic improvement of the Purcell 
factor of the cavity by a factor close to a thousand. Furthermore, as we 
will demonstrate in the last part of the paper, one can extremely decrease the 
mode volume of this cavity by simply shrinking accordingly the diameter of 
the wires and the period of the medium.


\begin{figure}[!h]
\begin{center}
\includegraphics[width=1\columnwidth]{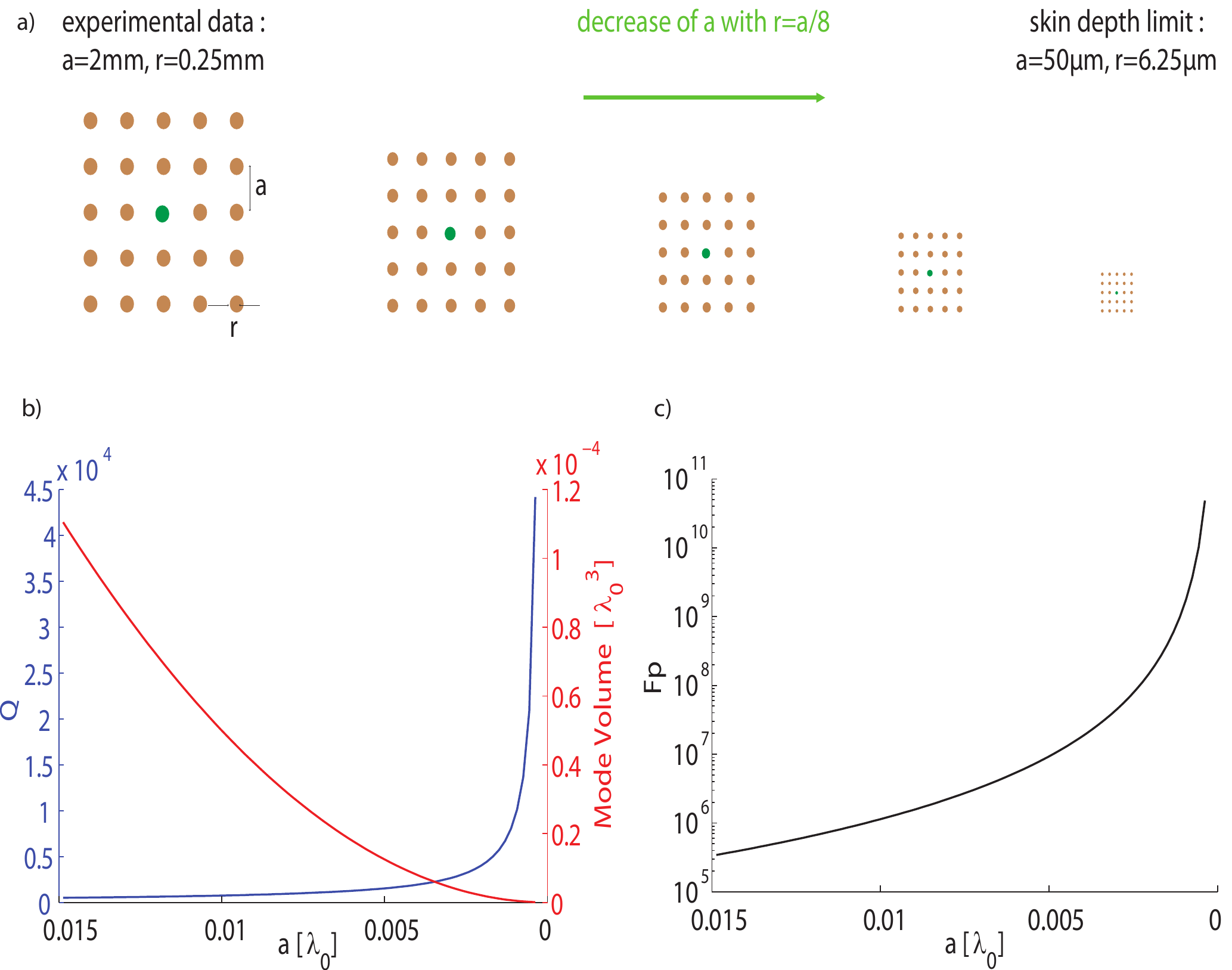}
\caption{Evolution of (b) the Q factor (blue) and mode volume $V_m$ (red), and (c) the Purcell factor when reducing the lattice parameter $a$ (and consequently the wire's radius)}
\label{Fig.4}
\end{center}
\end{figure}

Because we could not easily fabricate denser samples using thinner wires, we
present theoretical estimations of structures similar to the measured periodic one,
albeit with varying periods and diameters of wires. We fix the
ratio between the diameter of the wires and their period to 4, as in our
experimental ordered sample. We evaluate the consequences of reducing the diameter
of a wire from 0.5 mm down to a few skin depths (Fig 4.a). First, we can estimate
the resonance Q factor of such a cavity. As stated before, the
hybridization band gap is really efficient and forbids any radiation loss,
thus the cavity damping is given by losses only. Those are ohmic losses in
metals and they result from a penetration of the fields in the defect wire and
its direct neighbors. For each of them, the volume where the field
penetrates can be approximated as a skin depth wide ring of height $L$.
Hence we easily estimate that the amount of energy dissipated per cycle
decreases linearly with respect to the wires' radius since the ring of
field penetration decreases accordingly. While shrinking the dimensions,
we can therefore write $Q(r)/Q(r_0)=r_0/r$, $r_0$ and $Q_0$ being
respectively the initial radius and quality factor.
Second, the mode volume displays a quadratic evolution with respect to the
lattice constant (Fig 4.b). Indeed, the field is confined around the
defect in the $S=(2a)^2$ area formed by its nearest neighbors. Thus the
mode volume is estimated to be $V_m=\lambda/2S=2\lambda a^2=36 \lambda
r^2$, which increases quadratically with the wires' radius since we chose
to linearly vary it with the lattice parameter.
  
Hence, very strikingly, reducing the wires' radius and the period of the 
medium results at the same time in higher Q factors and lower mode volumes which is 
a very unusual behavior for cavities. Indeed, usually one has to choose 
between high Q factors\cite{Painter1999} or low mode volumes\cite{Oulton2009}. Here, the Purcell factor 
increases as $1/r^{3}$, solely limited by the skin depth in copper (Fig 4.c).


In summary, we have demonstrated ultra small mode volume defect cavities
in deep subwavelength periodic and random metamaterials made out of resonant electric wires. This proves
that the hybridization band gap of these metamaterials does not rely on spatial order.
We experimentally obtained cavities presenting quality factors around 500 and mode
volumes of about $\lambda^{\mathrm{3}}$/10000. We underlined that these structures can
be shrunk in the transverse dimensions down to an area of some skin depth
squared, hence providing at the same time incredibly small mode volumes and high quality factors
owing to a mitigation of ohmic losses. This results in enormous Purcell factors 
in the microwave domain. Similar results are expected in the THz range and the 
deep IR, since in those range noble metals still possess good conducting properties
\cite{Ordal:83}. In the visible, losses associated with plasmons should seriously
alter the quality factors mentioned here, but the concept could still lead to 
ultra small mode volumes cavities\cite{Yao12072011}. 
This type of cavities, because of their very high Purcell factors, 
should find applications in areas that rely on wave/matter interactions and 
nonlinear effects for instance in quantum-electrodynamics, energy rooting 
and filtering, or wave detectors and sources. Moreover, associated to the waveguides, 
 splitters and other discrete components that we demonstrated recently\cite{NatPhys}, they should lead to many applications. We can now imagine molding at wish the flow of waves in metamaterials not only
at a scale independent of the wavelength, but also in spatially random samples, as opposed to
photonic crystal demonstrations, which simplify fabrication issues.

\bibliography{biblio}

\end{document}